\documentclass[amsmath,eqsecnum,preprint,pre,aps,showpacs]{revtex4}
\usepackage{amsmath}
\usepackage{amsfonts}
\usepackage{graphicx}
\usepackage{amssymb}
\usepackage{dcolumn}
\usepackage{color}
\usepackage{ulem}
\usepackage{bm}   
\usepackage{float}
\begin{document}

\emph{}
\title{Induced phase transformation in ionizable colloidal nanoparticles}
\author{Leticia L\'opez-Flores$^{1}$ and Monica Olvera de la Cruz$^{1,2,3}$}

\affiliation{
$^{1}$Department of Materials Science and Engineering, Northwestern University, Evanston, 60208, Illinois, United States.\\
$^{2}$Department of Chemistry, Northwestern University, Evanston, 60208, Illinois, United States.\\
$^{3}$Department of Physics and Astronomy, Northwestern University, Evanston, 60208, Illinois, United States.
}

\date{\today}

\begin{abstract}
Acid-base equilibria directly \textcolor{black}{influence} the functionality and behavior of particles in a system. Due to the ionizing effects of acid-base functional groups, particles will undergo charge exchange. The degree of ionization and their intermolecular and electrostatic interactions \textcolor{black}{are} controlled by varying the pH and salt concentration of the solution in a system. Although the pH can be tuned in experiments, it is hard to model this effect using simulations or theoretical approaches. This is due to the difficulty in treating charge regulation and capturing the cooperative effects in a colloidal suspension with Coulombic interaction.  In this work, we analyze a suspension of ionizable colloidal particles via Brownian simulations and derive a phase diagram of the system as a function of pH. It is observed that as pH increases, particles functionalized with acid groups change their arrangement from face-centered cubic (FCC) packing to a disordered state. We attribute these transitions to an increase in \textcolor{black}{the} degree of charge polydispersity arising from an increase in pH. Our work shows that charge regulation leads to amorphous solids in colloids when the mean nanoparticle charge is sufficiently high.

\end{abstract}

\keywords{Charge regulation, phase transition, Crystallization, glass trasnition}

\maketitle

\vspace*{15em}
 To Fyl Pincus, with admiration for his groundbreaking contributions to charged colloids and polyelectrolytes that have inspired us, the Soft-matter community celebrates his enduring legacy and ongoing influence on our understanding of complex systems.
\newpage

\section{Introduction}
Ionizable surface functional groups are observed in biology and soft materials \cite{bookPodgornik}. These groups can release or combine with a hydrogen ion, and in this way acquire or lose charge, respectively \cite{Atkins}. Ionizable functional groups exist in numerous systems, such as polyelectrolytes, micelles, fibers, proteins, membranes, and colloids \cite{Szleiferpolyele,MarioTagliazucchi,Podgornik,Szleifer,Prusty,Leung,levin1,Szleifer2,Borkovec,Borkovec1,Erik}. An important property of these systems is their responsiveness to an increasing pH. An illustrative example of this phenomenon involves the reaction-driven release of ions from the surface of the particles into the surrounding solution \cite{Revbiomolecular,crmoleculas}. This process imparts cooperativity in the distribution of surface charge on the particles. That is, the charge on the particles modify the electrostatic interactions, which, in turn, modify the particles charges.  This recursive relation leads to the emergence of different states, such as glasses. Herein, we analyze how the charge distribution in a colloid suspension evolves and adapts to new chemical environments.

The process of charging and discharging an acid or base is referred to \textcolor{black}{as}  {\it charge regulation} (CR) \cite{Shumaker,Parsegian}. Charge regulation is a mechanism that maintains chemical equilibrium in complex systems in response to changes in pH, temperature, and ionic strength. Different approaches have been used to describe CR effects including theoretical, \cite{Szleifer1} and more recently, computer simulations \cite{Holm,Levin,Levin0,Erikjcp}. Moreover, these studies offer insights into the interplay between charge distribution and the behavior of charged entities in complex systems.  Because of the  practical implications in fields like nanotechnology, engineering, and pharmaceuticals, elucidating the phenomena that can control CR is of utmost importance.

The intersection of charge regulation and structural transitions introduces a challenge when electrostatic interactions play a central role. The CR effect reflects the adjustment of charge on pH responsive groups due to specific chemical reactions. This process profoundly influences the interaction of particles and molecules in a system, and significantly impacts the final structure. In particular, the readjustment of charge in a system can gradually slow the movement of particles, leading to a diverse landscape of structures that the system can access, such as crystalline or amorphous solids. Moreover, studying the interplay between CR and phase transitions presents a challenge due to the combination of chemical reactions and electrostatic interactions arising in these systems.

The objective of this paper is to elucidate the phase behavior in response to pH changes in a system of surface functionalized nanoparticles with acid-base groups.  Our model is analogous to nanoparticles or micelles with surface ionizable groups that can be observed in nature. The pH causes a change in the charge distribution on the surface of the nanoparticle, which imparts charge polydispersity leading to  different structures such as disordered, crystalline or fluid phases. To construct the phase diagram, we study the structural and dynamic properties as a function of pH by varying the density of nanoparticles in the system.

\section{Methodological Aspects}
\noindent We study the CR effect between negatively charged colloidal nanoparticles of diameter $\sigma_b=5l_B$, immersed in a uniform implicit solvent of dielectric constant $\epsilon$ and temperature $T$, where $l_B=e^2/(4\pi \epsilon_0 \epsilon k_B T)$ is the Bjerrum length, $e$ is the elementary charge, $k_B$ is the Boltzmann constant, and $\epsilon_0$ is the vacuum permittivity. Each nanoparticle contains $n_A$ weak acid groups carrying zero charge (neutral state $A$) or an elementary charge $e$ (dissociated state $A^-$), and is accompanied with dissociated ions ($H^+$, $OH^-$). The acid dissociation reaction is given by
\begin{equation}
A =A^- + H^+
\end{equation}
with equilibrium dissociation constant $pKa$. The nanoparticles are inmersed in an aqueous solution where $l_B=0.7$ nm, with typical pH and monovalent salt concentration values. We set  $pI_{S^{\pm}}=2$, which corresponds to 10 mM of salt; $n_A=110$ acid dissociation groups are represented by small spheres of diameter $\sigma=l_B$, and they are distributed on the nanoparticle's surface with a $pK_a=6.5$. The total number density of the nanoparticles,  $\rho$, is related to the volume fraction, $\phi=\pi \rho \sigma_c /6$. The excluded volume interaction is modeled by the Lennard-Jones (LJ) potential

\begin{equation}
u_{ij}(r)=\epsilon \Big[\Big( \frac{\sigma}{r}\Big)^{12}-4\Big( \frac{\sigma}{r}\Big)^6+1 \Big]\Theta(\sigma-r)
\label{hspotencial}
\end{equation}

where $\theta(r)$ is the unit step function and $\sigma$ is the diameter of the small particles. 

\begin{figure}[ht]
\centering
\includegraphics[scale=.35]{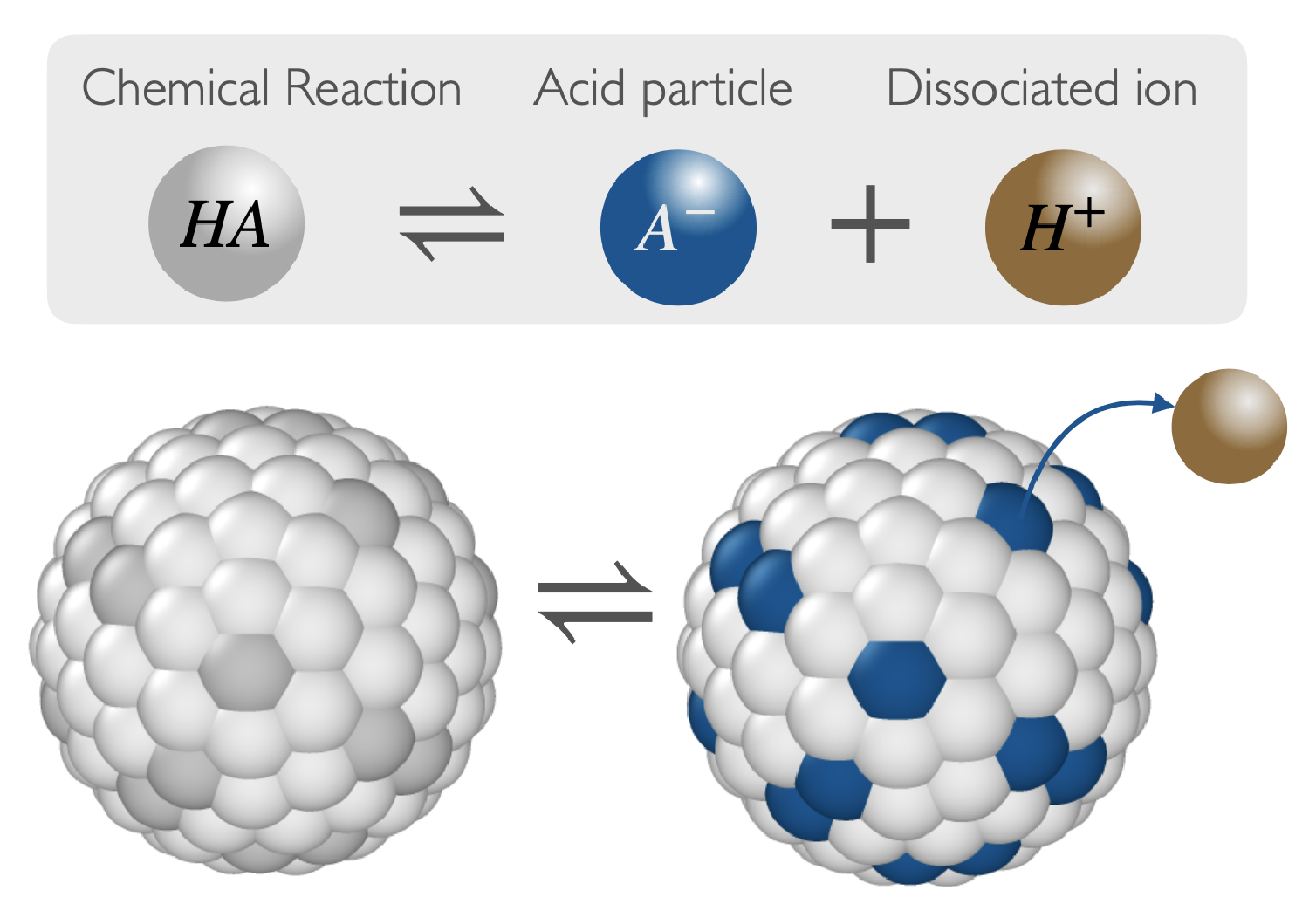}
\caption{Schematic model for a spherical nanoparticle under a chemical reaction $A \rightleftharpoons A^{-} + H^{+} $. The spherical particle carries the acid particles and delivers the dissociate ions to the solvent.}
\label{chem_reaction}
\end{figure}

The nanoparticles interact via LJ and Coulomb potentials.  We use the particle-particle-particle-mesh (PPPM) algorithm to solve the long-range electrostatic interactions. To compute the charge regulation effect, we use the methodology  \textcolor{black}{described} in the Ref. \cite{Erikjcp}. The temperature is controlled by a Langevin thermostat with a damping factor equal to 20$\tau$, where the unit time $\tau$  is based on the mass $m$ of the ions and dissociable groups. When the system is diluted, the time of equilibration is longer than when we have a concentrated system. Following equilibration, we obtain the particle charge average as a function of the pH for different concentrations of nanoparticles with fixed salt concentration. At low concentration, we recover the behavior reported in Ref.\cite{Erik} and the charge of one particle obtained using the Poisson-Boltzmann equation\cite{Pincus}.
The results are expressed in LJ units, where $M$, $\sigma$ and $\epsilon$ are the units of mass, length and energy, respectively. The time unit is $\tau=\sqrt{M\sigma^2/k_BT}$. The simulations were conducted with $N=108$ nanoparticles formed by $n_A$ small particles in a cubic simulation box with periodic boundary conditions. The initial configuration was placed on an FCC lattice at the desired density. We run the simulation for $10^5$ time steps, and the interaction between particles is only given by excluded volume interactions. This procedure is performed to generate a random configuration for all systems. Once the initial configuration is constructed, several thousand cycles are performed to allow the system to equilibrate with respect to CR of the nanoparticles and the addition of the ions and dissociable groups. This is followed by at least two million cycles and the data is collected every hundred cycles. The simulation starts with an electrically neutral system, where the particle is uncharged.  We compare a system with uncharged nanoparticles and a system with charged nanoparticles when subjected to different pH environments and find that both converge to similar charge distributions. Throughout the simulation, pair of ions are inserted or deleted to maintain the electroneutrality.  To compute an average charge distribution on the nanoparticles in our system we average over 100 different configurations. 

The structural and dynamic properties are calculated from the equilibrium configurations generated in the simulations. We follow the center of mass trajectory  of each nanoparticle to calculate the properties reported in this paper. Moreover, the radial distribution function was obtained by using the standard approach \cite{allen}. 

The mean squared displacement (MSD) is calculated to follow the transition from the liquid to a solid state. The MSD is given by 
\begin{equation}
W(t)=\langle [\Delta \vec{r}(t)]^2\rangle/ 6
\end{equation}
where $\Delta \vec{r_j}(t)=\vec{r_j}(t)-\vec{r_j}(0)$. In the limit of long time, the MSD reads $W(t)/\sigma^2 \approx  D_L^* t/\tau$, which is proportional to the scaled long-time self-diffusion coefficient $D_L^*\equiv D_L/D^0$. 

\section{pH-dependent phase transitions of systems with ionizable groups}

In this section, we analyze the simulation results for negatively charged ionizable groups on the surface of the nanoparticles as a function of pH. For this, we consider a liquid formed by $N=108$ nanoparticles in a volume V, interacting through an excluded-volume LJ potential, given by the Eq. (\ref{hspotencial}), where the diameter is given by the center of mass of the nanoparticle, i.e., $\sigma_b$. In addition, Coulomb interaction between all charges are computed.  The state space of the system is spanned by two variables: the number density $n=N/V$ and the pH. For fixed temperature $T=1$, the equilibrium phase diagram of the system is provided in the space (pH, $\phi$), which contains the crystalline and amorphous solid phases. 

A crystalline transition occurs as the density of the particles increases; the system undergoes a transition from a fluid-like state to an ordered crystalline state.  This transition is characterized by the sudden appearance of long-range positional order, where particles occupy specific positions within a repeating lattice. Instead, unlike crystallization, which involves the formation of an ordered lattice, the glass transition leads to the formation of a disordered state, which does not have long range translational order, hereafter refered to as an amorphous solid.

\begin{figure}[ht]
\centering
\includegraphics[scale=.3]{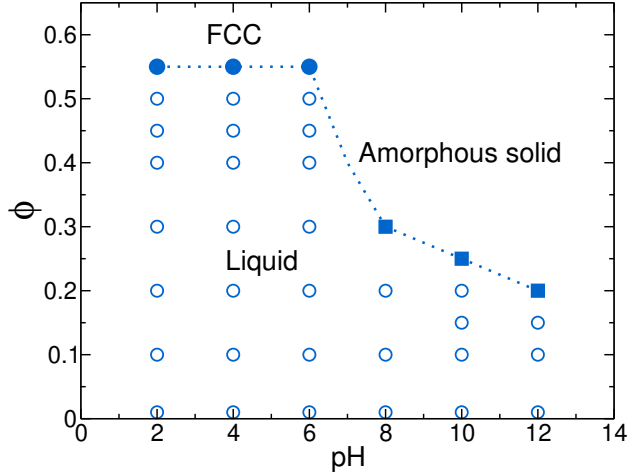}		
\caption{The state space of the system spanned by the variables ($\phi$, pH). The open circles imply the system is in a liquid state, and the full symbols imply the system has transformed into a different structure. For a pH between 2-6, the system is in a FCC structures. For a pH between 7-12, the structure is an amorphous solid.}
\label{diagram}
\end{figure}

From the analysis of structural and long-time asymptotic properties, we show in Fig. \ref{diagram} the liquid-crystalline-glass phase diagram as a function of pH. At low pH ($=2,4,6$), we observe when the density of colloidal nanoparticles is increased, the system forms an FCC crystalline structure. This phenomenon occurs when the volume fraction $\phi \simeq 0.55$. Nevertheless, if the pH exceeds 6, we observe the FCC structure is destroyed, and we get a disordered structure. To characterize these phases in our system, we analyzed the average charge of the particles, the configuration of the systems, and the static and dynamic properties. The analysis of all the phase transitions are presented below.

\begin{figure}[ht]
\centering
\includegraphics[scale=.3]{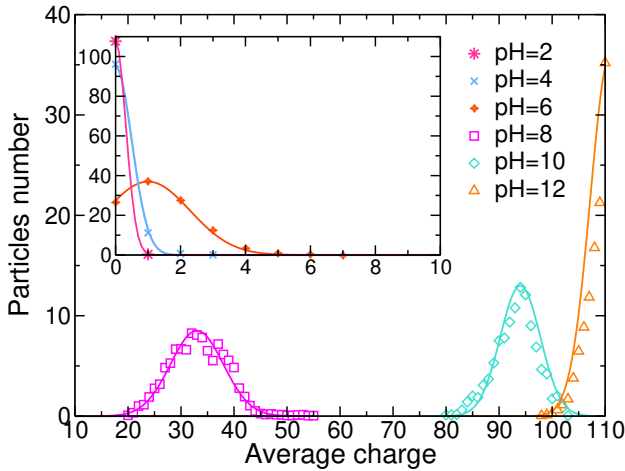}		
\caption{Average surface charge on the nanoparticle as a function of the number of total nanoparticles for different values of pH. The line represents a fitted Gaussian distribution given by the Eq. $f(x)=ae^{-c(x-b)^2}$.}
\label{carga_particula}
\end{figure}

Furthermore, the simulation shows that the nanoparticles are entirely discharged at low pH (pH=2), exhibiting only hard-sphere behavior. However, when pH is  \textcolor{black}{increased} to 4-6, we find that some surface groups are ionized, which allows us to get charged colloidal nanoparticles. In Fig. \ref{carga_particula}, we show the distribution of particle number with respect to average surface charge at different pH values. The average charge is obtained at the phase boundary of Fig. \ref{diagram}.  At pH=2 and 4, we find that each nanoparticle contains zero or a few charges on average on the surface when the system is in equilibrium. This result implies a narrow Gaussian distribution of charge on the surface of the nanoparticles.  \textcolor{black}{When} pH is increased to 6, we find that  \textcolor{black}{average} charge is normally distributed. However, the average charge of each nanoparticle in the system is around one ionizable ion on each particle, as shown in the inset of Fig. \ref{carga_particula}. This average charge is insufficient to break the FCC structure. To get disordered states, pH must be increased, the increase of the average charge on the nanoparticles. In the main panel of Fig. \ref{carga_particula}, we observe the average charge is a Gaussian distribution, and the average charge increases as a function of pH.

\begin{figure*}[ht]
\centering
\includegraphics[scale=.17]{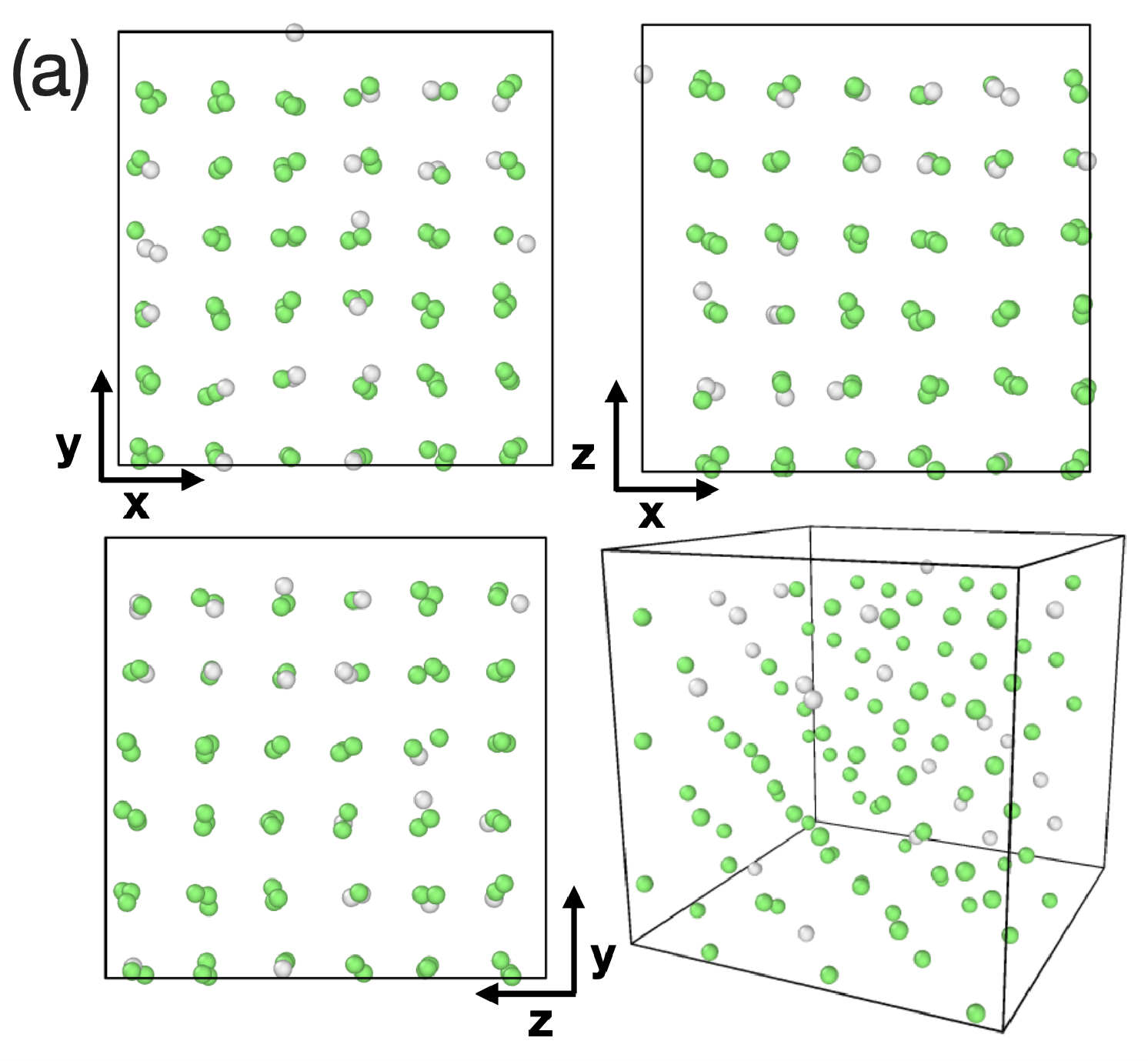}
\includegraphics[scale=.17]{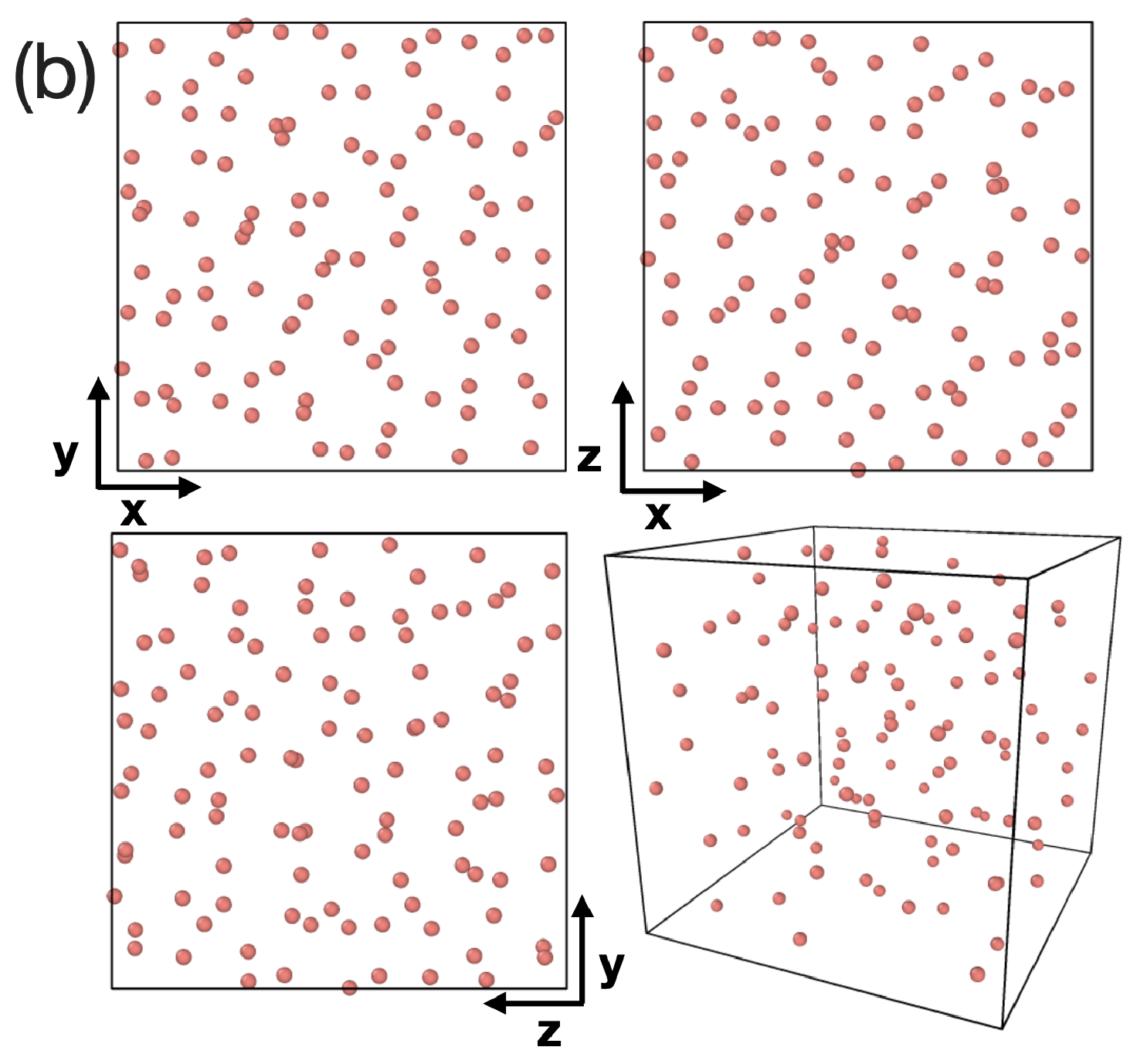}
\includegraphics[scale=.17]{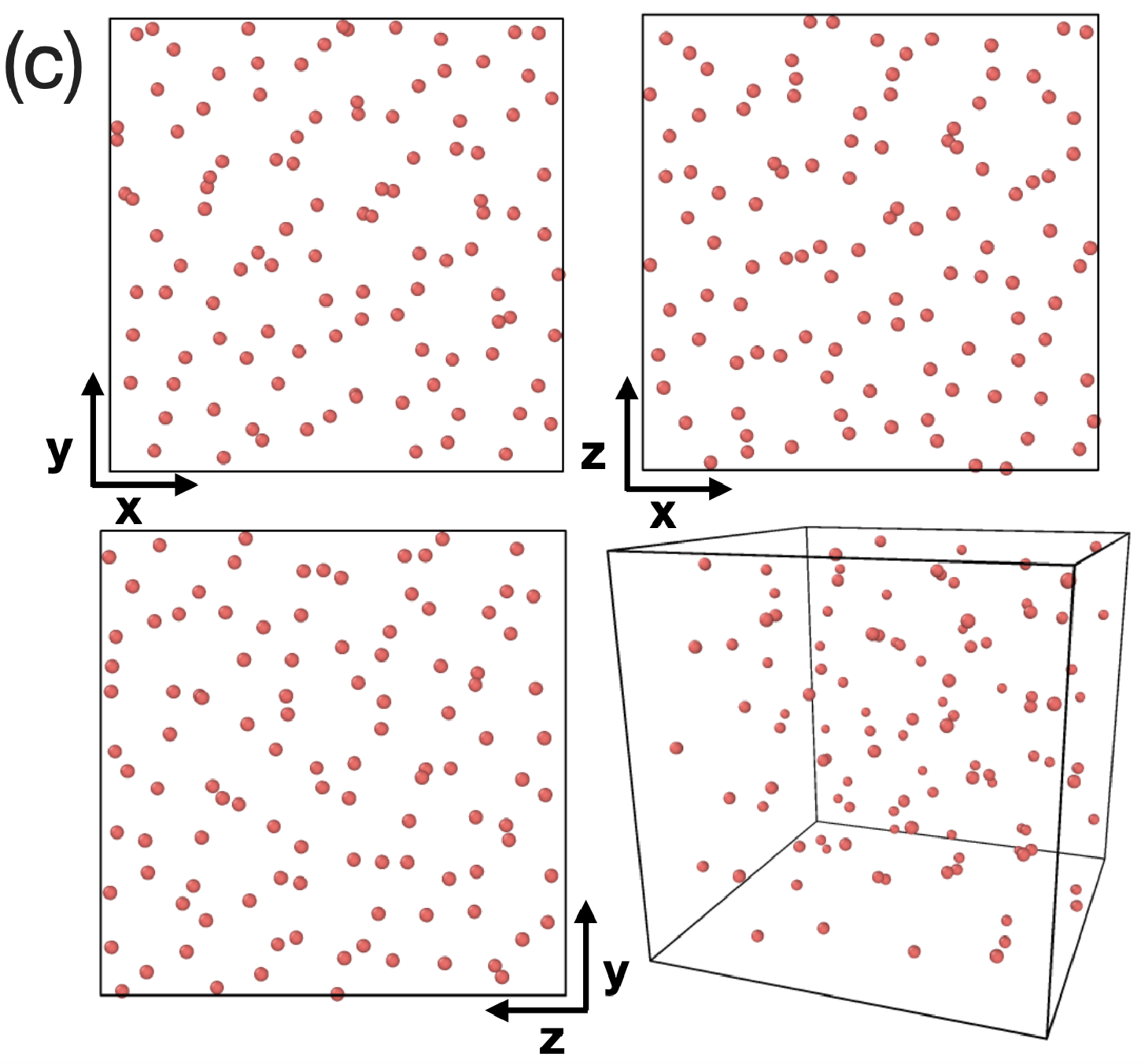}
\caption{Configuration of the center of mass of the nanoparticles at (a) (pH,$\phi$)=(2,0.55) , (b) (pH,$\phi$)=(8,0.30),  and (c) (pH,$\phi$)=(12,0.20). We show different perspectives of the system to determine the structure in the system.  }
\label{estructuras}
\end{figure*}

In Fig. \ref{estructuras}(a), we show the center of mass (CM) of the colloidal nanoparticle at pH=2. The configuration shows the CM of the nanoparticle maintained in an FCC structure, and the CM is around each order position in the structure. All the CM of the nanoparticles are around the position of the FCC structure. The phase diagram in Fig. \ref{diagram} shows that as pH exceeds 6, we observe a density decay. This decay is due to the presence of electrostatic interaction in the system. As mentioned previously, the surface charge of the nanoparticles increases as a function of the pH. This surface charge increase enhances the repulsion between particles as the density  \textcolor{black}{decreases}. These configurations are in Fig. \ref{estructuras}(b) and (c) where we can observe the CM of the nanoparticles \textcolor{black}{in} a fluid. The analysis of the static and dynamic properties provide information about the transition of the systems.

\begin{figure*}[ht]
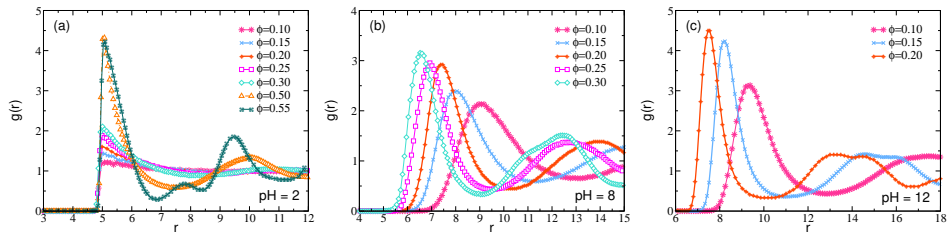

\centering
\includegraphics[scale=.15]{Figure5a.eps}
\includegraphics[scale=.15]{Figure5b.eps}
\includegraphics[scale=.15]{Figure5c.eps}
\caption{Radial distribution function of the center of mass of the nanoparticles at different values of $\phi$. The Fig. (a) provide the information of the system at pH=2, (b) at pH=6 and (c) pH=12.}
\label{gr_ph}
\end{figure*}

In the case of the static structure, we obtain the radial distribution function $g(r)$ at different concentrations and pH values. In Fig. \ref{gr_ph}, we plot simulation results for $g(r)$ of the CM of the nanoparticles at different pH. In Fig. \ref{gr_ph}(a), we show the result for pH=2, where at high densities $\phi=0.55$ we get an FCC structure. At low volume fraction $\phi$, we observe that $g(r)$ is a flat line around the value 1. This means the particles do not interact between them, i.e., the only interaction between them is excluded-volume. When the density is increased, we observe that $g(r)$ oscillates around 1, which means the particles are correlated. Also, we observe at $\phi=0.55$ we find the second peak at $r=7.5\sigma$, which means that the first and second nearest-neighbors are at one distance  equal to 1.54. If we have an FCC structure in a unit cell, the distance between the particle from its nearest-neighbors is when $r=\sqrt{2}=1.41\sigma$, which is close to the distance we get for the average of the $g(r)$. This information corroborates the result of the system is an FCC structure, obtained in Fig. \ref{estructuras}. However, if we change the pH to a higher value of 6, we observe that the peak around $r=1.7$ disappears, and the second peak now \textcolor{black}{occurs} at bigger distances. Additionally, the oscillations on $g(r)$ are more pronounced \textcolor{black}{at} lower volume fractions when compared with results obtained for pH between 2-6. This means the electrostatic interaction plays an important role when the pH exceeds 6. In this case, we need to analyze the dynamic properties to get information  \textcolor{black}{on} if the system is in a liquid state or is in some glassy state.

\begin{figure*}[ht]
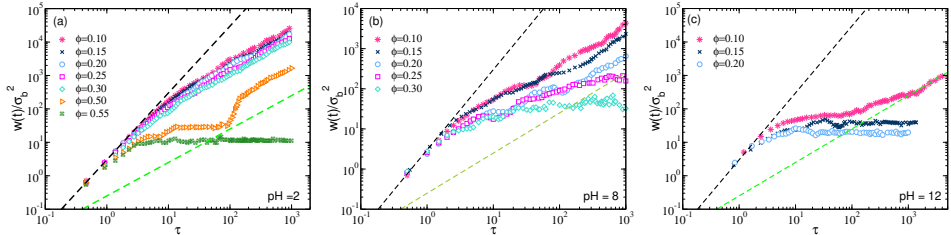

\centering
\includegraphics[scale=.15]{Figure6a.eps}
\includegraphics[scale=.15]{Figure6b.eps}
\includegraphics[scale=.15]{Figure6c.eps}
\caption{Mean square displacement for macromolecules at different volume fractions. The figure shows the MSD for different values of pH. These values are (a) pH=2, (b) pH=8 and (c) ph=12. The dark dashed line represents the quadratic behavior given by the expression $f(t)=3t^2$, and the green dashed line represents the linear behavior given by $f(t)=0.25t$}
\label{msd}
\end{figure*}

In Fig. (\ref{msd}), we show the MSD, following the center of mass of the nanoparticles, scale in  \textcolor{black}{reduced} units, i.e. $W(t)/\sigma_b^2$, which is a function of time scaled $t/\tau$. As we observe, the results display the ballistic behavior (at short times) [$W(t)\approx 3v_0^2t^2$] and a diffusive regime (at long times) [$W(t)\approx D_L t$] at different values of $pH$. Futhermore, we observe at low pH, the MSD has a flat behavior at $\phi=0.55$. This means that the particles are localized for a long time. However, when we have a disordered system, we observe a flat behavior when we increase the volume fraction. This is a well-known effect when we have disordered states or glasses\cite{Poon,Makse,Weitz}. This means that the  \textcolor{black}{particles} are localized for a long time. That is, a particle is caged by the neighbors, demostrating a glass phase \cite{Cipelletti,Laurati}.  Using all these elements, we characterize the phase diagram as a function of the pH for a system with ionizable groups on the nanoparticles.

\section{Conclusion}\label{sec13}

We analyze charge regulation effects on colloidal solutions of nanoparticles that contain acid-ionizable groups on their surface via simulations. The simulations include Coulomb interactions and hard-core repulsions. Charge regulation is an entropic effect that affects the charge of the particles, and the Coulomb potential energy affects their interactions, which, in turn, modify the charge on the nanoparticles. The phase diagram as a function of the pH is obtained by analyzing the structure and dynamical properties, as well as the configurations of the equilibrium states. The phase diagram shows a liquid state at any pH value when the volume fraction of nanoparticles is low. When the pH value is low and the nanoparticles have low or nearly zero charge, a crystalline structure emerges at large nanoparticle volume fractions. Instead, at large pH values such that the electrostatic interactions are significant, a disordered structure arises at intermedium nanoparticle volume fractions. We find that the disordered states have a large degree of charge polydispersity; that is, when the average nanoparticle charge is sufficiently high, charge regulation allows for large charge fluctuations. Interestingly, size polydispersity in charged colloids \cite{Hempelmann,LuisE,pedro,Nohely} and neutral nanoparticles with hard-core repulsions \cite{vanMegen,Tartaglia,Henderson,MCT, Fuchs,Leticia,zacarelli,gabriel} leads to glassy states. Here, charge polydispersity when Coulomb interactions are large leads to glassy states.  This work opens  \textcolor{black}{a} large number of questions for further discussion, such as the effect of large concentration of salt, multivalent ions, degree of charge polydispersity in binary solutions of particles that can take charge values of opposite sign or in macromolecules adsorbed to oppositely charged surfaces. Further understanding of charge regulation and phase transitions in complex systems, will allow us to engineer and tailor materials with enhanced properties and/or functionalities.  

\section*{Acknowledgments}
The authors acknowledge Felipe Jimenez Angeles and Brandon Onusaitis for helpful discussions and advised comments. This work was supported by the Sherman Fairchild Foundation and U.S. Department of Energy (DOE), Office of Science, Office of Basic Energy Sciences, under Award No. DE-FG02-08ER46539. 

\section*{Author contribution statement}
MOdlC supervised research; LLF performed research;  LLF and MODlC designed research, analyzed data and wrote the paper.

\section*{Availability of data and materials}
The datasets generated and/or analyzed during the current study are available from the corresponding author on reasonable request.

\section*{Conflict of interest} The authors declare no conflict of interests.

\end{document}